# Collective biphoton temporal waveform of photon-pair generated from Doppler-broadened atomic ensemble

Heewoo Kim[1], Hansol Jeong[1], and Han Seb Moon[1,2,*]

[1]*Department of Physics, Pusan National University, Geumjeong-Gu, Busan 46241, Republic of Korea*
[2]*Quantum Sensors Research Center, Pusan National University, Geumjeong-Gu, Busan 46241, Republic of Korea*
*[*]hsmoon@pusan.ac.kr*

Photonic quantum states generated from atomic ensembles will play important roles in future quantum networks and long-distance quantum communication because their advantages, such as universal identity and narrow spectral bandwidth, are essential for quantum nodes and quantum repeaters based on atomic ensembles. In this study of the biphoton temporal waveform (BTW) of the photon pairs generated from a cascade-type two-photon-transition, we report the collectively coherent superposition of biphoton wavefunction emitted from different velocity classes in a Doppler-broadened cascade-type atomic ensemble. We experimentally demonstrate that the three times difference of temporal width of both BTWs varies dependent on the wavelengths of the signal and idler photons from both $6S_{1/2}$–$6P_{3/2}$–$6D_{5/2}$ and –$8S_{1/2}$ transitions of $^{133}$Cs, corresponding to the idler and signal wavelengths of 852 nm–917 nm and 852 nm–795 nm, respectively. Our results help understand the characteristics of biphoton sources from a warm atomic ensemble and can be applied to long-distance quantum networks and practical quantum repeaters based on atom–photon interactions.

Quantum light sources generated from atomic ensembles are remarkable as crucial quantum resources for implementing photonic quantum information technologies, such as quantum memory and quantum repeaters, for quantum networks because the optical frequency and bandwidth of such photons are well suited for atom–photon interactions [1-5]. Bright and narrowband quantum light sources generated from atomic ensembles have been developed in various atomic systems [6-25]. Such photon sources based on atoms remain an unresolved challenge for the development of practical quantum sources. However, despite the use of an atomic medium, the experimental setups of atomic ensemble photon sources using atomic vapor cells are easy to use, simple, compact, robust, and inexpensive. In addition, they are comparable to those of conventional quantum sources for quantum optics obtained via the spontaneous parametric down-conversion (SPDC) process from nonlinear crystals [19]. Furthermore, the photons obtained via the spontaneous four-wave mixing (SFWM) process from warm atomic ensembles are excellent quantum resources for practical quantum photonic sources in terms of brightness, spectral bandwidth, and signal-to-noise ratio. Such SFWM photon pairs generated from warm atomic vapors have been successfully demonstrated in key quantum optics experiments, including time-resolved Hong–Ou–Mandel (HOM) interference, high-visibility Franson interference of time-energy entangled photon pairs, and entanglement swapping of photon-polarization qubits [26-31].

In particular, the important properties of narrow-band photon sources based on atomic ensembles have been characterized via measurements of the biphoton temporal waveform (BTW) because the coherence time of photons is longer than the time resolution of a single-photon detector (SPD) [32]. In the atom–photon interactions of a warm atomic ensemble, we should consider the inhomogeneous broadening due to the thermal motion of atoms with the Maxwell-Boltzmann distribution. Doppler broadening affects the properties of the photons generated from the warm atomic ensemble because the velocity class of the atoms satisfied by the two-photon resonance is determined in the warm atomic vapor [33]. The wavelengths of the interacting lasers and the emitted photons can affect the BTW of a photon pair from a moving atom with a different velocity because it presumably integrates the BTWs of the photon pairs emitted from each atom with the Maxwell–Boltzmann velocity distribution.

In the case of the cascade-type atomic system shown in Fig. 1(a), to minimize the Doppler effect in the two-photon resonance, the optimal spatial alignment of the counter-propagating geometry of the two contributing lasers was used for photon-pair generation from Doppler-distributed moving atoms [17]. However, in the cascade-type energy diagram of real atoms, two-photon Doppler broadening occurs because of the wavelength difference between the two lasers used for two-photon excitation. Furthermore, the wavelength ratio of the two lasers can significantly affect the two-photon coherence in a Doppler-broadening atomic system. Therefore, photon-pair generation from warm atomic ensembles may consider various phenomena due to the velocity distribution of warm atoms because the Doppler effect affects the two-photon coherence and the transitions between hyperfine states.

In this study, we report the collective effects of BTWs of photon-pair generation via SFWM processes in both cases of the cascade-type $6S_{1/2}$–$6P_{3/2}$–$6D_{5/2}$ and –$8S_{1/2}$ transitions of the warm $^{133}$Cs atomic ensemble, as shown in Fig. 1(b). In particular, the selected two-photon transitions in Fig. 1(b) are similar in terms of the wavelength difference but different in terms of the wavelength ratio. In the simple cascade-type atomic model shown in Fig. 1(a), we theoretically investigated the second-order cross-correlation function between the signal and idler photons, based on the wavelength ratio of the pump and coupling lasers. We discuss the coherent superposition of biphoton wavefunctions from different velocity classes in inhomogeneous, broadened atomic media. We compare the characteristics of the photon pairs in both cases and analyze the tendency of the second-order cross-correlation function based on the wavelength ratio from the theoretical results. These results provide a deep understanding of the interesting collective effect of the BTW from warm atomic ensembles with velocity distributions and are expected to contribute to the development of single-photon sources based on atomic vapor cells.

theoretical model involves the $6S_{1/2}(F=4)$–$6P_{3/2}(F'=5)$–$6D_{5/2}(F''=6)$ and –$8S_{1/2}(F''=4)$ transitions of $^{133}$Cs, as shown in Fig. 1(b). In these two-photon transitions, the wavelengths ($\lambda_{C2}$ and $\lambda_{C1}$) of the coupling lasers correspond to 917 nm and 795 nm, respectively. The natural linewidths of the $6D_{5/2}$ and $8S_{1/2}$ states are 2.6 MHz and 1.7 MHz, respectively [34]. The wavelength differences $|\lambda_P - \lambda_C|$ of these transitions are calculated to be 65 nm and 57 nm, and the $\lambda_C : \lambda_P$ ratios are estimated to be 1.08 and 0.93, respectively.

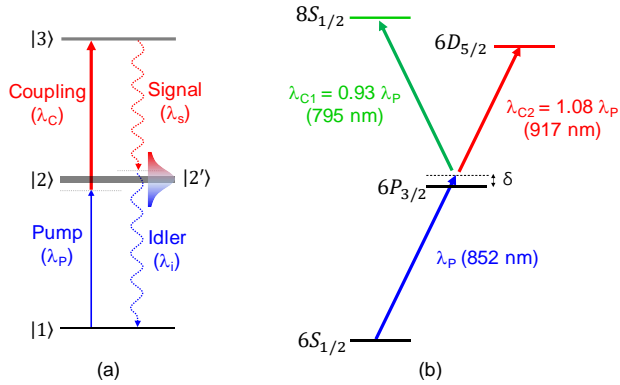

**Figure 1. Spontaneous four-wave mixing (SFWM) process in a cascade-type atomic system.** (a) Configuration for the SFWM process of a three-level cascade-type atomic system in the presence of pump and coupling lasers. (b) Cascade-type atomic systems of the $6S_{1/2}(F=4)$–$6P_{3/2}(F'=5)$–$6D_{5/2}(F''=6)$ and –$8S_{1/2}(F''=4)$ transitions of the $^{133}$Cs atom.

Here, the cascade-type configuration is defined so that each wavelength of the signal and idler photons is the same as that of the pump and coupling lasers for two-photon coherence. To evaluate the two-photon wavefunction for photon pairs via SFWM in a Doppler-broadened cascade-type atomic ensemble system, Fig. 1(a) shows the energy diagram of a three-level cascade-type atomic system in the presence of the pump and coupling lasers with wavelengths of $\lambda_P$ and $\lambda_C$, respectively. This cascade-type atomic system consists of a ground state $|1\rangle$, an intermediate state $|2\rangle$ ($|2'\rangle$ for spontaneous decay channel), and an excited state $|3\rangle$.

In this work, the cascade-type atomic system used in the

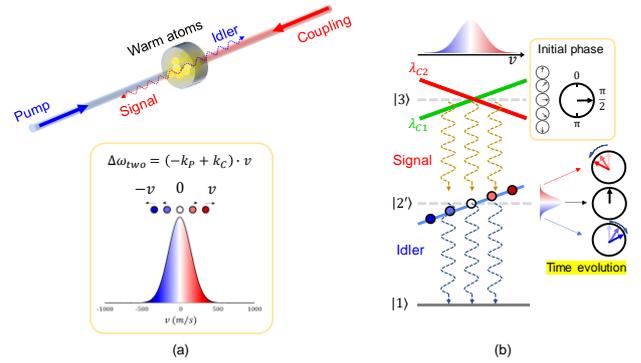

**Figure 2. Doppler effects in a cascade-type atomic system.** (a) Two-photon Doppler shift ($\omega_{two}$) from the warm atomic ensemble interacting with the counter-propagating pump and coupling lasers. (b) Biphoton wavefunction according to different velocity classes in the Doppler-broadened atomic media.

Using the counter-propagating geometry of the two contributing lasers in the cascade-type configuration, as shown in Fig. 2(a), photon pairs are generated in the phase-matched direction from the Doppler-distributed moving atoms. The temporal correlation between the signal and idler photons can be collectively enhanced by the two-photon coherence that interacts with the pump and coupling lasers in a cascade-type atomic system [27].

The two-photon Doppler shift ($\omega_{two}$) for the count-propagating pump and coupling lasers can be expressed as $\omega_{two} = (-k_p + k_C) \cdot v$, where $k_p$ and $k_C$ are the wave vectors of the pump and coupling lasers, respectively, and $v$ is the atom velocity. In the case of $\lambda_P > \lambda_C$ for the $6S_{1/2}$–$8S_{1/2}$ transition, indicated by the green line of the $|3\rangle$ excited state in Fig. 2(b), $\omega_{two}$ is positively increased relative to the moving atoms of the pump laser's propagating direction. In contrast, in the case of $\lambda_P < \lambda_C$ for the $6S_{1/2}$–$6D_{5/2}$ transition (red line), $\omega_{two}$ is negatively increased, opposing the direction of atom velocity.

The second-order cross-correlation function $g_{SI}^{(2)}(\tau)$ for the paired photon generated from a Doppler-broadened atomic ensemble via SFWM is defined as [35, 36]

$$g_{SI}^{(2)}(\tau) = \left| \int \Psi_v(\tau) f(v) dv \right|^2, \quad (1)$$

Here, $\Psi_v(\tau) \equiv \langle 0 | \hat{a}_i(t_i) \hat{a}_s(t_s) | \Psi_v \rangle$ is the two-photon wavefunction for the SFWM process in a Doppler-broadened cascade-type atomic system, where $t_s$ and $t_i$ denote the detection times of the signal and idler photons, respectively, and $\tau = t_i - t_s$ is the time difference between the signal and idler photon detection. $\Psi_v(\tau) = A_v e^{i\Phi_v}$ can be expressed as a complex function with the amplitude $A_v$ and phase $\Phi_v$ of the biphoton wavefunction. $f(v)$ is a one-dimensional Maxwell-Boltzmann velocity distribution function.

We can describe $g_{SI}^{(2)}(\tau)$, integrated with the velocity $v$ of atoms with a one-dimensional Maxwell–Boltzmann velocity distribution along the propagation direction of the pump beam, as follows (see Supplemental Material [37]):

$$g_{SI}^{(2)}(\tau) = \left| \int_{-v}^{v} dv \frac{1}{\sqrt{\pi}u} e^{-(\frac{v}{u})^2} \frac{C}{[4\{\frac{\Gamma_{21}}{2} + i(\delta_P - k_P v)\}\{\frac{\Gamma_{32} + \Gamma_{32'}}{2} + i(-k_P + k_C)v\} + \Omega_C^2]} e^{ik_I v\tau} e^{-\frac{\Gamma_{2'1}}{2}\tau} H(\tau) \right|^2 \quad (2)$$

where $C$ denotes the coefficient related to the transition rate. $\Gamma_{2'1}$ denotes the the decay rate of the idler photon mode; δ denotes the detuning; k denotes the wave vector; Ω denotes the Rabi frequency; τ denotes the time difference between signal and idler photon detection; and $u$ denotes the most probable velocity of a warm atomic ensemble. The subscripts denote the energy levels and the pump (P) and coupling (C) lasers.

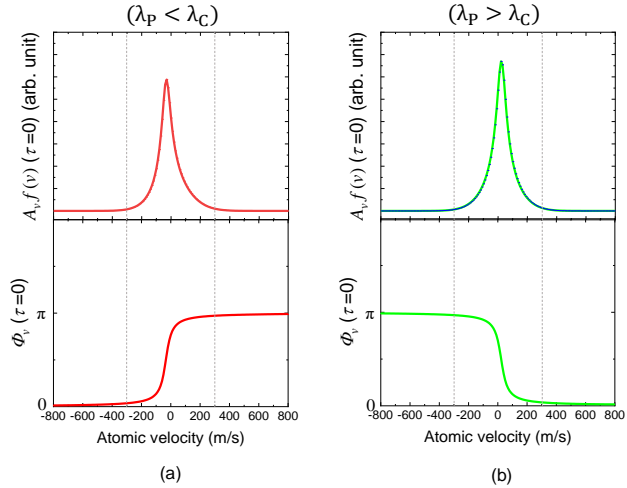

**Figure 3. Phase and amplitude of BTW from the Doppler-broadened atomic ensemble.** (a) Calculated amplitude $A_v f(v)$ and phase $\Phi_v$ of the $\Psi_v(\tau = 0)$ as a function of atomic velocity in two cases of (a) $\lambda_P < \lambda_C$ and (b) $\lambda_P > \lambda_C$, respectively.

To elucidate the cause of the wavelength dependence on the BTW of the photon pair generated from the Doppler-broadened atomic ensemble, we investigated the two-photon wavefunction from different velocity classes and the cross-correlation functions in inhomogeneous broadened atomic media. Figures 3(a) and 3(b) show the calculated results for the amplitude $A_v f(v)$ and phase $\Phi_v$ of $\Psi_v(\tau)$ in the cases of $\lambda_P < \lambda_C$ and $\lambda_P > \lambda_C$.

First, the amplitude $A_v f(v)$ as a function of the velocity of a warm atomic ensemble showed which component of the velocity groups contributed to the generation of photon pairs in both cases. Compared with the calculated amplitudes, although the height and center position of $A_v f(v)$ differ slightly in both cases, the shapes of both $A_v f(v)$ are similar because the wavelength differences $|\lambda_P - \lambda_C|$ of these transitions are similar.

Second, we compared with the phases of $\Phi_v$ as a function of the atomic velocity. $\Phi_v$ depends on the decay rate, detuning frequency, wave vectors, and time difference (τ) between the signal and idler photon detection. The total phases in both cases were out of phase with each other at $\tau = 0$. As shown in the initial phase ($\Phi_v$) of Fig. 2(b) at $\tau = 0$, when the phase of the zero-velocity atoms (grey dashed line of $|3\rangle$) is $\pi/2$, the positive value of $\omega_{two}$ is close to zero, but the negative value is close to $\pi$. This implies that the phase $\Phi_v$ of the biphoton wavefunction is changed according to each atom with a Maxwell–Boltzmann velocity distribution.

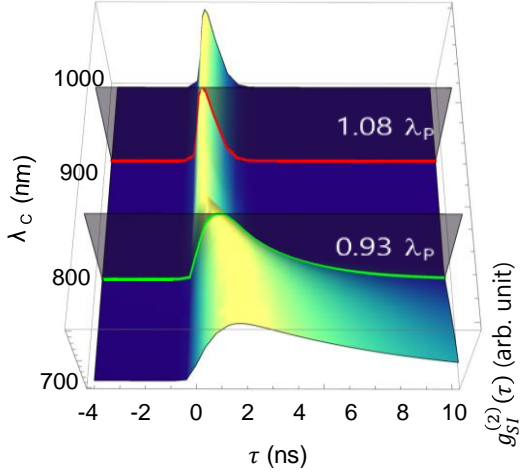

**Figure 4. Simulation of $g_{SI}^{(2)}(\tau)$ as a function of $\lambda_C$.** Results obtained by varying $\lambda_C$ from 700 nm to 1000 nm using coefficients of the 6S$_{1/2}$(F=4)–6P$_{3/2}$(F'=5)–8S$_{1/2}$(F''=4) transition at $\lambda_P$ = 852 nm. The red and green curves correspond to the 6S$_{1/2}$(F=4)–6P$_{3/2}$(F'=5)–6D$_{5/2}$(F''=6) and 6S$_{1/2}$(F=4)–6P$_{3/2}$(F'=5)–8S$_{1/2}$(F''=4) transitions, respectively. $g_{SI}^{(2)}(\tau)$ accounts for detector timing jitter modeled in a Gaussian function with a FWHM of 0.35 ns and the maximum value adjusted to 1.

To investigate the wavelength dependence on the photon-pair generation from the Doppler-broadened atomic system, we calculated $g_{SI}^{(2)}(\tau)$ as a function of the coupling wavelength $\lambda_C$. Figure 4 shows the simulation of $g_{SI}^{(2)}(\tau)$ in the range of $\lambda_C$ from 700 nm to 1000 nm, under the condition of $\lambda_P$ = 852 nm. Under experimental circumstances, the timing jitter of a SPD and the emitting order between the signal and idler photons should be considered. The time uncertainty of the SPD used in our experiment is 0.35 ns, including the electronic timing jitter. The timing jitter of the SPDs is assumed to be a Gaussian function. The practical cross-correlation function $g_{SI}^{(2)}(\tau)$ can be expressed as the convolution integral.

We can calculate the $g_{SI}^{(2)}(\tau)$ spectrum of the 6S$_{1/2}$(F=4)–6P$_{3/2}$(F'=5)–6D$_{5/2}$(F''=6) transition from Eq. (2), where the $\Psi_v(\tau)f(v)$ component is coherently superposed owing to the collective two-photon coherence effect, as indicated by the red curve in Fig. 4. The temporal width of the $g_{SI}^{(2)}(\tau)$ spectrum was calculated to be approximately 1 ns. In the case of $\lambda_P > \lambda_C$, the $g_{SI}^{(2)}(\tau)$ spectrum of the 6S$_{1/2}$(F=4)–6P$_{3/2}$(F'=5)–8S$_{1/2}$(F''=4) transition was in the range of $\tau > 0$ and close to the calculated $g_{SI}^{(2)}(\tau)$ spectrum, as shown by the green curve in Fig. 4. The main cause of the wavelength dependence on the $g_{SI}^{(2)}(\tau)$ spectrum generated from the Doppler-broadened atomic ensemble is the collective two-photon coherence effect, which is the coherent superposition of a two-photon wavefunction according to the atomic velocity.

Therefore, the spectral width of the BTW is related to the correlation time of the signal and idler photons, which are the key characteristics of the heralded single photons. In view of the BTW of the photon pair generated from the Doppler-broadened atomic ensemble, the theoretical result in Fig. 4 is interesting because the important property of the photon pair is changed by only the wavelength ratio of the pump and the coupling lasers. Furthermore, if our photon pair is time-frequency entangled, the spectral biphoton waveform of the photon pair may be changed by the wavelength ratio of both lasers under the same conditions, such as transition rates, decay rates, Doppler-broadening, and Rabi frequencies.

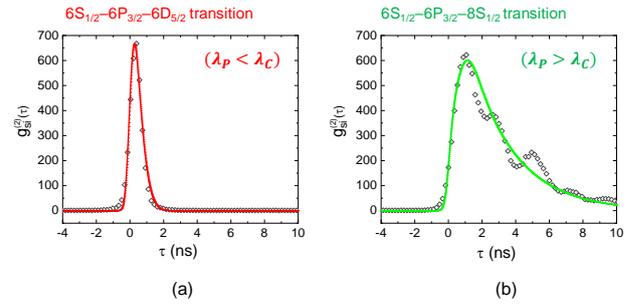

**Figure 5. Comparison with temporal biphoton waveforms (BTWs).** (a) The experiment results (black circles) and the theoretical result (red curve) of the cross-correlation function of $g_{SI}^{(2)}(\tau)$ in the 6S$_{1/2}$(F=4)–6P$_{3/2}$(F'=5)–6D$_{5/2}$(F''=6) transition. (b) The experiment results (black circles) and the theoretical result (greed curve) of the cross-correlation function of $g_{SI}^{(2)}(\tau)$ in the 6S$_{1/2}$(F=4)–6P$_{3/2}$(F'=5)–8S$_{1/2}$(F''=4) transition.

In our system, the two-photon wavefunction $\Psi_v(\tau)$ in Eq. (1) is coherently superposed in an atomic ensemble with a Maxwell-Boltzmann distribution because of the collective two-photon coherence of the Doppler-broadened atomic ensemble with the two-photon resonant interaction between the pump and coupling lasers. However, when the wavelengths of the pump and coupling lasers differed, the two-photon Doppler shift remained in a two-photon transition. The partial component ($\Delta v_{two}$) of the velocity groups in the Doppler-broadened atomic ensemble contributed to the generation of two-photon coherence and can be expressed as

$$\Delta v_{two} = \frac{\Gamma_3}{k_P - k_C}, \quad (3)$$

where $\Gamma_3$ denotes the natural linewidth of the $|3\rangle$ excited state. Considering the natural linewidths of the $6D_{5/2}$ and $8S_{1/2}$ states, the photon pairs obtained via the two-photon transition contribute to the velocity-selective classes of atoms in the warm atomic vapor, corresponding to a $\Delta v_{two}$ of 31.3 m/s and 20.2 m/s and contributing to the SFWM process in the Doppler-broadened atomic ensemble.

In the transitions shown in Fig. 1(b), we experimentally demonstrated photon-pair generation from the chip-scale Cs vapor cell and measured the cross-correlation functions under the experimental parameters of the 10 μW pump and 5 mW coupling lasers. The Cs vapor cell used in our experiment had chip-scale vapor cell outer dimensions of 2 mm × 3.5 mm × 1.4 mm [38]. The small space for photon-pair generation had a diameter of 5 mm and a thickness of 1 mm. The temperature of the vapor cell for photon pair generation was set to 85 °C, and the optical depth was estimated to be ~10. As shown in Fig. 2(a), the counter-propagating pump and coupling lasers spatially overlap in the vapor cell. Under the phase-matching condition of the counter-propagating direction of the photon pair, the successfully generated signal and idler photons from the chip-scale Cs vapor cell were coupled into two single-mode fibers positioned at a tilt angle of ~2° relative to the propagating directions of the pump and coupling lasers, respectively.

We measured the BTWs of photon pairs in both transitions of a warm Cs atomic ensemble. Figure 5 shows the second-order cross-correlation functions from the measured coincidence histogram and the normalized cross-correlation function $g_{SI}^{(2)}(\tau)$ as a function of the time delay ($\tau$) from the signal to the idler photons, normalized to the accidental coincidence count, for the two cases of the $6S_{1/2}(F=4)$–$6D_{5/2}(F''=6)$ and $6S_{1/2}(F=4)$–$8S_{1/2}(F''=4)$ transitions. We measured the coincidence counts for a 180 s effective measurement time using a time-correlated single-photon counter. The BTWs of narrow-band photon sources based on atomic ensembles can be measured because the coherence time of the photons emitted from the atomic ensemble is longer than the timing jitter of the SPD (Excelitas SPCM-AQRH-13).

In Fig. 5(a), the black circles show the measured $g_{SI}^{(2)}(\tau)$ spectrum with a high signal-to-noise ratio (SNR), which was estimated to be the maximum value of $g_{SI}^{(2)}(\tau)$ (approximately 667). The red solid curve in the case of $\lambda_P < \lambda_C$ indicates the theoretical result, including an SPD timing jittering of 0.35 ns. The experimental results were in good agreement with the theoretical results corresponding to the calculated curve at $\lambda_C = 1.08\lambda_P$. Here, the temporal width of $g_{SI}^{(2)}(\tau)$ is estimated to be 1.0(1) ns, which is approximately two times smaller than the inverse of the Doppler-broadening linewidth of the warm $^{133}$Cs atoms.

However, in the case of $\lambda_P > \lambda_C$ in Fig. 5(b), the temporal width of $g_{SI}^{(2)}(\tau)$ is estimated to be 2.8 ns, which is three times larger than that in the case of $\lambda_P < \lambda_C$. In the experimental results, the oscillation feature of the $g_{SI}^{(2)}(\tau)$ spectrum denoted the quantum beats of the BTW of the photon pair generated via SFWM multi-channels relating to the hyperfine states of the intermediate state of $6P_{3/2}$ [38]. The envelope of the measured $g_{SI}^{(2)}(\tau)$ spectrum, except for the oscillation due to quantum beats, is in close agreement with the green solid curve in Fig. 5(b), which corresponds to the calculated curve at $\lambda_C = 0.93\lambda_P$. Therefore, by comparing the two $g_{SI}^{(2)}(\tau)$ spectra of the photon pair emitted from different transitions in the single warm atomic vapor cell, we confirmed the wavelength dependence of the collective BTW in a Doppler-broadened atomic ensemble for the first time.

In conclusion, we have demonstrated that the collective two-photon coherence in a Doppler-broadened cascade-type atomic ensemble affects the BTW of time-frequency entangled biphotons. Using the single chip-scale Cs vapor cell, we successfully generated high-SNR photon pairs in both the cascade-type $6S_{1/2}$–$6P_{3/2}$–$6D_{5/2}$ and –$8S_{1/2}$ transitions. For the first time, we could find the wavelength dependence of the photon pair from the Doppler-broadened cascade-type atomic ensemble generation via SFWM processes. By analyzing the amplitude and phase terms of the calculated biphoton wavefunction, we determined that the main cause of the wavelength dependence on the BTWs of the photon pair is the coherent superposition of the biphoton wavefunction from different velocity classes in the Doppler-broadened atomic ensemble. We confirmed that the BTW of the photon pair depends significantly on the wavelengths of the signal and idler photons emitted from two cascade-type two-photon transitions counter propagation of the pump and the coupling laser.

Although the photons emitted from the atomic medium offer the advantage of universal identity, the properties of the photon-pair sources from warm atomic ensembles can be significantly changed according to the wavelength conditions for atom–photon interactions. We hope that our findings can be applied to studies of quantum networks and quantum optics, including quantum-entanglement swapping between completely autonomous sources based on atomic vapor cells.

**Acknowledgment**

This study was supported by the National Research Foundation of Korea (NRF) (Grant No. NRF-2021R1A2B5B03002377 and RS-2023-002831466 2182065300101), and the Institute of Information & Communications Technology Planning & Evaluation (IITP) (IITP-2024-2020-0-01606 and Grant No. 2022-0-01029). Additionally, this research received support from the 'Regional Innovation Strategy (RIS)' through the NRF funded by the Ministry of Education (MOE) (2023RIS-007).

**Data availability**

The data that support the findings of this study are available from the corresponding author upon request.